\documentclass[prb,aps,amsmath,amssymb,amsfonts,floatfix,groupeaddress,showpacs,twocolumn]{revtex4-1}

\usepackage{amsmath}
\usepackage{amssymb}
\usepackage{xspace}
\usepackage{graphicx}
\usepackage{grffile}
\usepackage{nicefrac}
\usepackage{capt-of}
\usepackage{color}

\graphicspath{{figs/}}

\newcommand{\bk}{\mathbf{k}}

\begin{document}

\title{Low-Energy Model and Electron-Hole Doping Asymmetry of Single-Layer 
Ruddlesden-Popper Iridates}
\author{Alexander Hampel}
\affiliation{I. Institut f{\"u}r Theoretische Physik,
Universit{\"a}t Hamburg, D-20355 Hamburg, Germany}
\author{Christoph Piefke}
\affiliation{I. Institut f{\"u}r Theoretische Physik,
Universit{\"a}t Hamburg, D-20355 Hamburg, Germany}
\author{Frank Lechermann}
\affiliation{I. Institut f{\"u}r Theoretische Physik, Universit{\"a}t Hamburg, 
D-20355 Hamburg, Germany}

\begin{abstract}
We study the correlated electronic structure of single-layer iridates based on
structurally-undistorted Ba$_2$IrO$_4$. Starting from the first-principles
band structure, the interplay between local Coulomb interactions and 
spin-orbit coupling is investigated by means of rotational-invariant slave-boson 
mean-field theory. The evolution from a three-band description towards an
anisotropic one-band ($J$=$\nicefrac{1}{2}$) picture is traced. Single-site and 
cluster self-energies shed light on competing Slater- and
Mott-dominated correlation regimes. A nodal/anti-nodal Fermi-surface dichotomy
is revealed at strong coupling, with an asymmetry between electron and hole 
doping. Electron-doped iridates show clearer tendencies of Fermi-arc formation, 
reminiscent of hole-doped cuprates.

\end{abstract}

\pacs{71.27.+a,71.18.+y,71.70.Ej,71.30.+h}

\maketitle

\section{Introduction}
Iridum oxides based on the Ruddlesden-Popper series pose a particular 
challenging electronic structure problem.~\cite{cra94,kim08,jac09} The cooperation of 
strong spin-orbit coupling (SOC) with $5d$-shell Coulomb interactions stabilizes 
insulating phases at stoichiometry below room temperature. Since these
compounds usually show also antiferromagnetic (AFM) ordering, it is debated if 
Mott- or Slater mechanisms rule the observed insulating 
states.~\cite{mar11,ari12,hsi12} Despite formally assumed weaker electronic 
correlations, the question arises if iridates still display deeper 
analogies to layered ruthenates or high-T$_c$ cuprates in view of non-BCS
superconducting properties.~\cite{wat13}
   
\begin{figure}[t]
\begin{center}
(a)\includegraphics*[height=5cm]{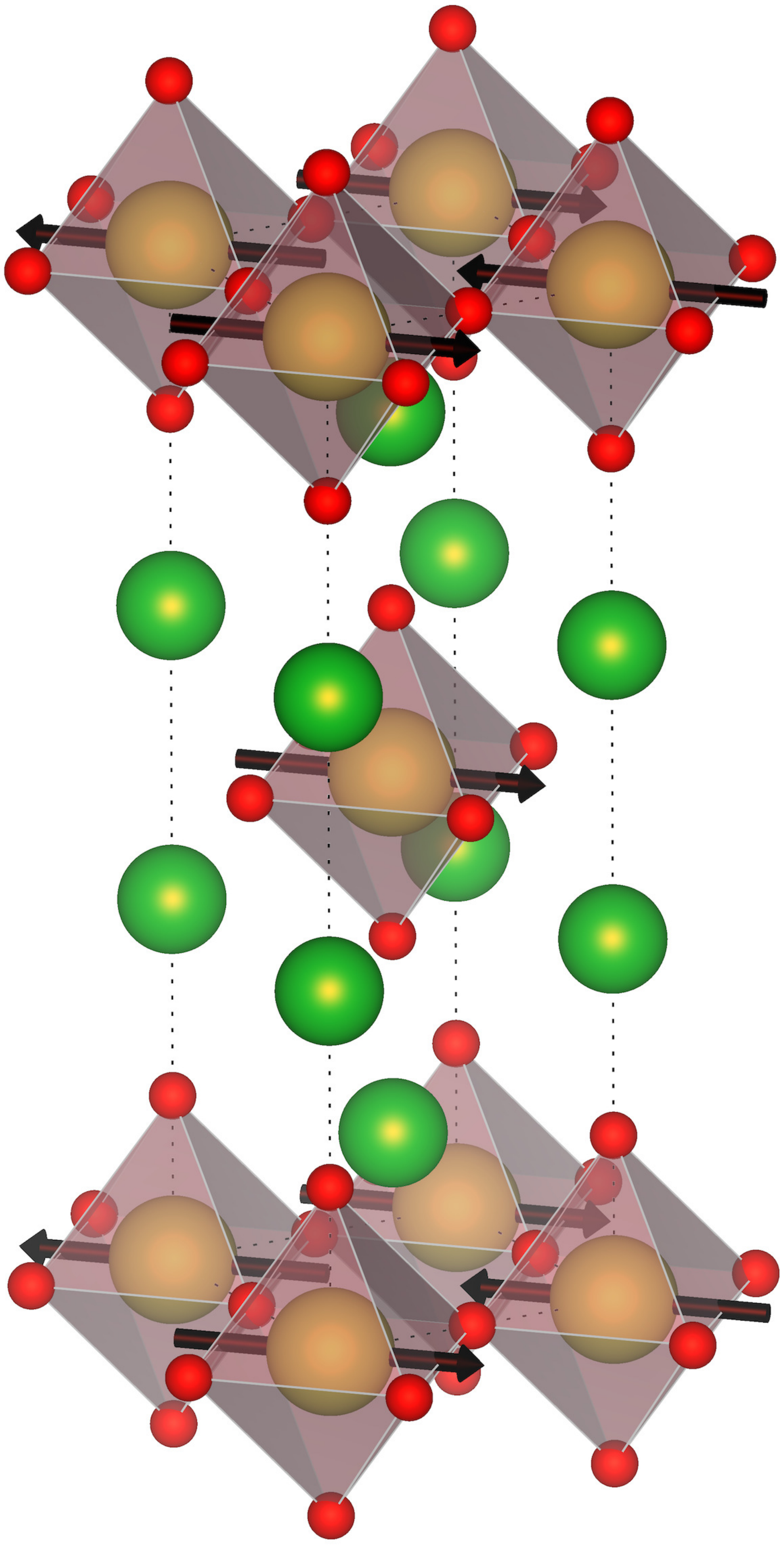}\hspace*{0.3cm}
\includegraphics*[height=4.5cm]{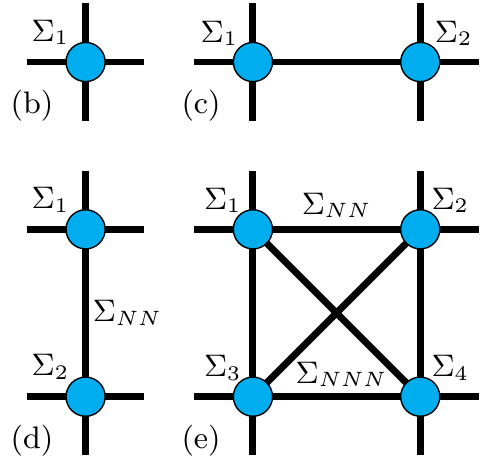}
\end{center}
\caption{(color online) (a) Crystal and AFM structure of tetragonal 
Ba$_2$IrO$_4$, with Ba (green), Ir (brown) and O (small red) ions. 
(b-e) Utilized self-energy representations within the square
lattice of an IrO$_2$ layer. (b) single on-site and (c) two on-site 
self-energies, neglecting inter-site terms. (d) NN two-site cluster 
and (e) four-site (2$\times$2) self-energy.
\label{fig:intro}}
\end{figure}

While the Sr compound of single-layer ruthenates has ideal tetragonal symmetry,
the sister compound Sr$_2$IrO$_4$ shows tilting of the IrO$_6$ octahedra. In contrast
Ba$_2$IrO$_4$ (see Fig.~\ref{fig:intro}a) is again free from 
distortions~\cite{oka11} and thus serves as a canonical system with a single Ir 
ion in the paramagnetic (PM) unit cell.~\cite{mos14,uch14} The AFM insulating phase of 
Ba$_2$IrO$_4$ has an Ir local magnetic moment of $0.34\mu_{\rm B}$, 
with an easy axis perpendicular to the $c$-axis~\cite{bos13}, and is stable up to 
$T_{\rm N}$=240K. Only a small charge gap of about $\sim$0.2 eV is deduced from 
angle-resolved photoemission spectroscopy (ARPES) measurements.~\cite{mos14}

Theoretical studies of (Ba,Sr)$_2$IrO$_4$ based on variational 
Monte-Carlo~\cite{wat10} 
as well as density functional theory (DFT) combined with dynamical mean-field 
theory~\cite{mar11,ari12,zha13} support the original heuristic picture of a 
correlation-mediated spin-orbit driven insulator. Therein the SOC discriminates the 
Ir $5d(t_{2g})$ into effective $J_{\rm eff}$=$\nicefrac{1}{2},\nicefrac{3}{2}$ 
states.~\cite{per14} While four electrons of Ir$^{4+}$ fill up  
$J_{\rm eff}$=$\nicefrac{3}{2}$ completely, one electron remains in 
$J_{\rm eff}$=$\nicefrac{1}{2}$ at low energy. The interacting half-filled band at 
the Fermi level is then either gapped mainly due to the Slater mechanism forming an 
AFM state or directly by electronic correlations with secondary magnetic ordering.

Doping of the iridates is achievable,~\cite{kor10} and recent experimental works 
succeded to reveal a subtle electronic structure for both electron- and 
hole doping.~\cite{kim14,cao14,he15,bro15,tor15,kim15} By surface electron doping of
Sr$_2$IrO$_4$,~\cite{kim14} the quasiparticle (QP) strength seems to vary along 
the Fermi surface, somehow reminiscent of the famous Fermi arcs known from 
hole-doped cuprates. Though effective hole-doping of Sr$_2$IrO$_4$ also shows
$k$-selective features,~\cite{cao14} but the fermiology appears much more 
incoherent.

In this work we focus on single-layer tetragonal Ba$_2$IrO$_4$ as a test 
case for basic accounts on the intriguing spin-orbit assisted correlation physics. 
From the realistic band structure at stoichiometry, effective low-energy three- and
one-band Hubbard models are constructed to assess the possible correlation regimes. 
Local and non-local self-energy representations are employed to study 
metal-insulator transition and doping effects. Fermi-surface differentiations in 
qualitative agreement with recent experimental findings are revealed. An obvious 
dichotomy in the doped fermiology between electron and hole doping is found at
strong coupling, identifying the electron-doped case as the candidate for a 
proper analogue to the hole-doped cuprates.

\section{Theoretical framework}
First-principles DFT calculations in the local density approximation (LDA) are 
performed for Ba$_2$IrO$_4$ in the $I4/mmm$ space group according to crystal data 
by Okabe {\sl et al.}.~\cite{oka11} Computations are performed using a mixed-basis 
pseudo-potential scheme~\cite{lou79,mbpp_code} with~\cite{hei10} and without the 
inclusion of spin-orbit coupling. We construct maximally-localized Wannier functions 
(MLWFs)~\cite{mar12}
for the Ir $5d(t_{2g})$-based low-energy bands close to the Fermi level from LDA 
calculations without SOC. Therefrom an initial three-band Hubbard Hamiltonian in 
the original $5d(t_{2g})$ basis of orbitals $m,m'$=$yz,xz,xy$ and with local 
spin-orbit term on Ir sites $i$ is drawn, i.e.
\begin{equation}
H=\hspace*{-0.1cm}\sum_{\bk mm'\sigma}\hspace*{-0.1cm}\varepsilon^{t_{2g}}_{\bk mm'}
c^{\dagger}_{\bk m\sigma}c^{\hfill}_{\bk m'\sigma}+
\sum_{i}\left(H^{(i)}_{\rm CF}+H^{(i)}_{\rm SO}+H^{(i)}_{\rm INT}\right)
\;,\label{eq:thrham}
\end{equation} 
where $c^{\dagger}$,$c^{\hfill}$ are creation, annihilation operators for the MLWF
states with spin projection $\sigma$=$\uparrow,\downarrow$. The $t_{2g}$ dispersion 
$\varepsilon^{t_{2g}}_\bk$ excludes on-site parts, which enter the
crystal-field term $H_{\rm CF}$. A Slater-Kanamori parametrization with 
Hubbard $U^{t_{2g}}$ and Hund's exchange 
$J^{t_{2g}}_{\rm H}$=$0.14$eV~\cite{kat14,per14} is used for $H_{\rm INT}$, 
including density-density as well as spin-flip and pair-hopping terms. The SO interaction
reads $H_{\rm SO}$=$\lambda\sum_\nu {\bf s}_\mu \cdot {\bf l}_\mu$, where $\lambda$ is 
the coupling constant and {\bf s},{\bf l} are spin-, angular-momentum operators. 
Because of the shift of $5d(e_g)$ to higher energies, restricting the general 
spin-orbit interaction matrix to the $5d(t_{2g})$ manifold is 
justified.~\cite{wat10,car13}

The full problem (\ref{eq:thrham}) is solved by mean-field rotational-invariant 
slave-boson (RISB) theory,~\cite{li89,bue98,lec07} using a multi-orbital single-site 
self-energy (see Fig~\ref{fig:intro}b) for the correlated subspace of three
effective $t_{2g}$ orbitals. The method amounts to a distinction of the electron's QP 
(fermionic $f_{\nu\sigma}$) and high-energy excitations (taken care of by the set
of local slave bosons $\{\phi\}$) on the operator level through 
$\underline{c}_{\nu\sigma}$=$\hat{R}[\{\phi\}]^{\sigma\sigma'}_{\nu\nu'}f_{\nu'\sigma'}$, 
where $\nu$ is a generic orbital/site index.~\cite{lec07} 
Self-energies with a term linear in frequency and a static part result in 
mean-field. The RISB approach is especially suited to 
model anisotropic interactions,~\cite{schu12} and here allows to treat the 
interacting spin-orbit problem in complete generality, i.e. without abandoning
off-diagonal terms. Neglecting $H_{\rm INT}$ leads to spin-orbit QP bands in very 
good agreement with the LDA+SOC low-energy dispersion.

For larger $\lambda$, the three-band Hamiltonian may be reduced to a tailored 
one-band problem for the effective $J$=$\nicefrac{1}{2}$ state at low-energy. In this 
restricted orbital space we also allow for an enlarged correlated subspace in 
real space via clusters of two and four sites (see Fig.~\ref{fig:intro}d,e). Therewith
non-local correlations up to next-nearest neigbor (NNN) are incorporated. The initial
cluster embedding is of cellular type, $k$-dependent self-energies are obtained
for the two-site ($\Sigma^{\rm (2)}$) cluster and the four-site 
($\Sigma^{\rm (4)}$) cluster via further periodisation using~\cite{bir04}
\begin{eqnarray}
\Sigma^{\rm (2)}(\bk,\omega)=&&\hspace*{0.2cm} \Sigma_{11}^{\rm (2)}(\omega) + 
\Sigma_{12}^{\rm (2)}(\omega)\, (\cos  k_x + \cos  k_y)\;,\\
\Sigma^{\rm (4)}(\bk,\omega)=&&\hspace*{0.2cm} \Sigma_{11}^{\rm (4)}(\omega) + 
\Sigma_{12}^{\rm (4)}(\omega)\, (\cos k_x + \cos k_y)\\ \nonumber
&&+\; \Sigma_{13}^{\rm (4)}(\omega) \cos k_x\,  \cos k_y\;.
\end{eqnarray}
Since single-site RISB is equivalent to single-site DMFT with a simplified impurity
solver, the cluster extension corresponds to cluster-DMFT with the named restrictions
in the self-energy representation. Albeit approximative, the cluster-RISB method has 
been proven capable to shed light onto relevant features of non-local correlation 
physics.~\cite{lec07,lec09,fer09_2,maz14}

\section{From three-band to effective one-band physics}
\begin{figure}[t]
\includegraphics*[width=8.5cm]{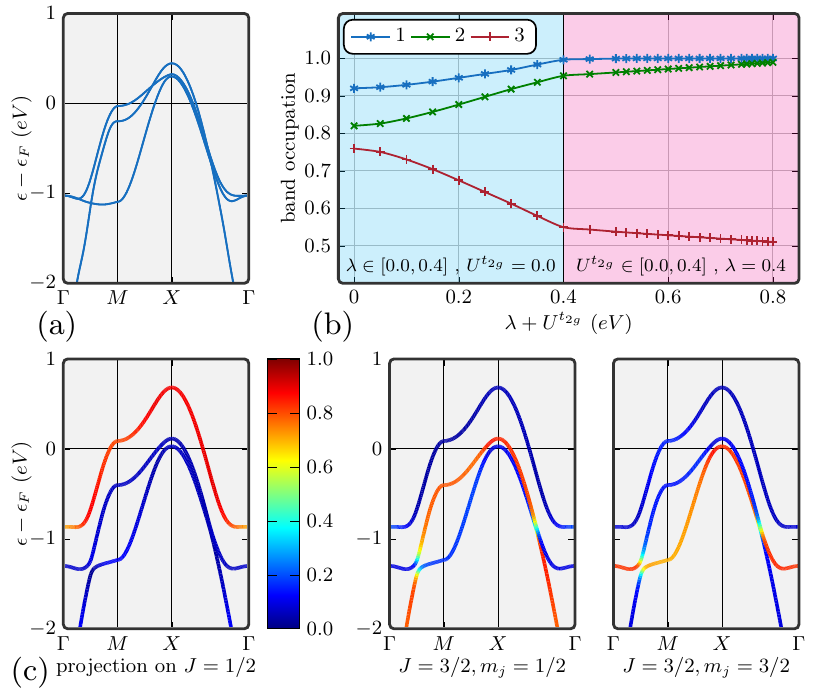}
\caption{(color online) Dispersion and occupations within the iridate $t_{2g}$
three-band manifold. (a) Bands without SOC, (b) band fillings with increasing SOC
strength $\lambda$ and Hubbard $U$ (with '3' denoting highest band). 
(c) effective-$J$ weight on the respective bands.
\label{fig:bandocc}}
\end{figure}

The LDA calculations for Ba$_2$IrO$_4$ reveal dominant $t_{2g}$-like bands at 
low-energy, and a minor $e_g$-like electron pocket around $\Gamma$.
Static DFT+U computations lead to an upward energy shift of the latter pocket into
the unoccupied region. Thus that $e_g$-derived contribution plays no
vital role in the key correlation physics and is neglected in the following.
Figure~\ref{fig:bandocc}a displays the MLWF-based $t_{2g}$-like low-energy bands 
adapted from LDA without SOC. Including spin-orbit coupling in the subsequent
RISB treatment shifts the lower band manifold with effective $J$=\nicefrac{3}{2}
down in energy (see Fig.~\ref{fig:bandocc}b). Inclusion of $H_{\rm INT}$ shifts 
those bands even further away from the Fermi level $\epsilon_{\rm F}$,
eventually resulting in completely filled $J$=\nicefrac{3}{2} and half-filled
$J$=\nicefrac{1}{2} states (cf. Fig.~\ref{fig:bandocc}c). This limit may be 
understood from a constructive interplay between Hund's third rule and the 
minimization of Coulomb interactions in the Ir$(5d^5)$ shell. 
The orbital character of the remaining half-filled band at $\epsilon_{\rm F}$ is 
indeed nearly exclusively of $J$=\nicefrac{1}{2} kind. Due to its isolation, 
the low-energy physics of single-layer iridate can be further analyzed
to a good approximation within a one-band picture. From the three-band calculation
with $\lambda$=0.4eV and neglecting $H_{\rm INT}$, we therefore Fourier transform the 
isolated $J$=\nicefrac{1}{2} band to obtain a single-band tight-binding 
parametrization. 
In addition to a local Coulomb interaction scaling with a Hubbard $U$, a 
nearest-neighbor (NN) spin-spin interaction 
term is introduced to take care of the spin-orbit induced in-plane 
$J$=\nicefrac{1}{2} pseudo-spin ordering.~\cite{jac09} The low-energy one-band 
iridate Hamiltonian is then given by
\begin{equation}
H_{\rm 1B}=\sum_{ij\sigma}t_{ij}c^{\dagger}_{i\sigma}c^{\hfill}_{j\sigma}+
U\sum_i n_{i\uparrow}n_{i\downarrow}+
\Gamma\sum_{\langle ij\rangle}S_i^{||}S_j^{||}
\;,\label{eq:1bham}
\end{equation}
where $t_{ij}$ marks the hoppings of the underlying $J$=\nicefrac{1}{2} dispersion 
with bandwidth $W$=1.55 eV and $\Gamma$$>$0 as the anisotropic AFM pseudo-spin 
coupling between the in-plane component $S^{||}$ of the pseudo-spins. The first 
near-neighbor in-plane hoppings amount to $(t,t',t'',t''')$=$(-205, -16, 35, 13)$ meV,
and the inter-layer coupling is given by $t_\perp$=$-$11meV. Based on the work of 
Katukuri {\sl et al.},~\cite{kat14} a value $\Gamma$=12meV is computed for 
the anisotropic interaction. Note that the effective one-band description does not
allow to discriminate between different ordering axes of the pseudo-spins, the 
definite in-plane easy axis remains arbitrary.~\cite{kat14}

Albeit in the following we focus on in-plane aspects, the complete three-dimensional 
dispersion is included for deriving the effective one-band physics within mean-field 
RISB. Half filling is generally marked by the effective one-orbital occupation $n$=1. 

\section{Effective one-band physics from single-site RISB}
\begin{figure}[t]
\includegraphics*[width=8.5cm]{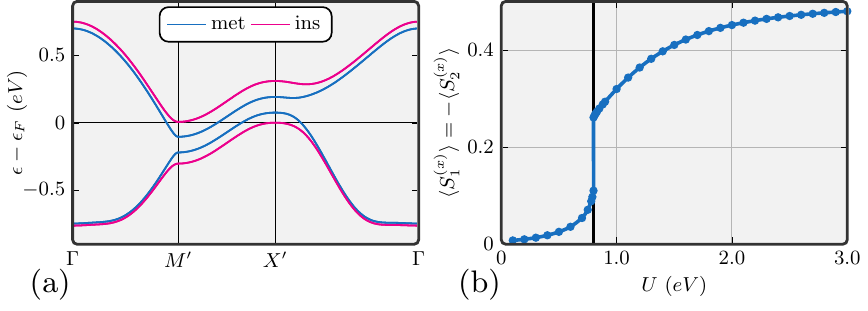}
\caption{(color online) Two-single-site MIT with AFM order
in the effective one-band model. (a) Metallic and insulating QP bandstructure for
$U$=0.8 eV. (b) Jump of $\langle S^{(x)}\rangle$ at the first-order MIT.
\label{fig:afmmit}}
\end{figure}

Lets first focus on the pure on-site self-energy treatments, neglecting inter-site 
terms. Disregarding the spin-spin interaction, the PM Mott transition with 
vanishing QP weight 
$Z$=$\left[1-\frac{\partial}{\partial\omega}\Sigma\,\right]_{\omega=0}^{-1}$
occurs at $U_{\rm c,PM}$=2.85 eV, i.e. $U_{\rm c,PM}/W$$\sim$1.84. To 
account for AFM order we use a $\sqrt{2}$$\times$$\sqrt{2}$ unit-cell 
architecture, treating two NN Ir ions with their respective on-site $\Sigma$
(cf. Fig~\ref{fig:intro}c). The anisotropic interaction between the pseudo-spins is 
chosen favorably along the $x$-direction and handled in mean-field decoupling, i.e.
$S_i^{(x)}S_j^{(x)}$$\rightarrow$$S_i^{(x)}\langle S_j^{(x)}\rangle$. At 
stoichiometry antiferromagnetism with staggered moments aligned along the $x$-axis 
marks the ground state for any $U$$>$0. 
For $U_{\rm c,AFM}$=0.8eV the system becomes insulating at a 
first-order transition (see Fig.~\ref{fig:afmmit}a). Thus the critical  
$U$ for the metal-insulator transition (MIT) is strongly lowered when allowing for 
magnetic order. Figure.~\ref{fig:afmmit}b shows that the spin moment pointing along
$x$ becomes highly susceptible to small interaction changes around 
$U_{\rm c,AFM}$, but saturates only at much larger interaction strength. Therefore
in this case the MIT is not of strong Mott type, i.e. does not result in complete 
electron localization. It has magnetic-driven signature, where the charge-gap opening
results in the formation of increased-dispersive Slater-like bands.~\cite{wat10}
Away from stoichiometry, the AFM order remains stable up to rather large doping,
as long as $\langle S^{(x)}\rangle$ is finite. Symmetric 30$\%$ electron/hole 
doping is necessary to render $\langle S^{(x)}\rangle$$\rightarrow$0 for $U$=1eV.

\section{Effective one-band physics from cluster RISB}
To evaluate the relevance of inter-site self-energy contributions especially in the 
doped regime, we extend the one-band investigations towards computations within a 
cluster framework. Therein the pseudo-spin interaction term in eq.~(\ref{eq:1bham}) 
may be treated in complete many-body form on the local clusters. It is directed along
$x$ for the case of a two-site cluster, and symmetrically along $x$,$y$ in the four-site
cluster approach.

In the following the analysis of the self-consistent statistical weight of 
cluster multiplets via the resulting slave-boson amplitudes $\{\phi\}$ will prove
useful. Note that the local cluster eigenstates can be written as
$|\underline{\Gamma}\rangle$$\sim$$\sum_{\Gamma'}\phi^{\hfill}_{\Gamma\Gamma'}|{\rm vac}\rangle|{\Gamma}'\rangle$, whereby $\Gamma,\Gamma'$ share
the same quantum numbers.~\cite{lec07,fer09_2} Here the eigenbasis is labelled
by the set ${\cal B}$=$\{N,S^2,S_z,(H_{\rm loc})\}$, with $N$ as the total
particle number, $S^2$ the total spin momentum, $S_z$ its $z$-component and 
$H_{\rm loc}$ as the local energy. If in the following $H_{\rm loc}$ breaks spin
symmetry, the resulting states are treated in a first-order perturbation approach.

There are 16 eigenstates on the two-site cluster and 256 on the four-site cluster.
The statistical weight of states $\Gamma_q$ with identical quantum numbers 
according to ${\cal B}$ is collected in the probability 
\begin{equation}\label{statweight}
\rho_q=\sum_{q'}\rho_{qq'}\delta_{qq'}=\sum_{q'p}\phi^*_{pq}\phi^{\hfill}_{pq'}\delta_{qq'}\quad,
\end{equation}
with the normalization $\sum_p\rho_p$=${\rm Tr}\,(\phi^\dagger \phi)$=1.

\subsection{ Two-site cluster}
\begin{figure}[b]
\includegraphics*[width=8.5cm]{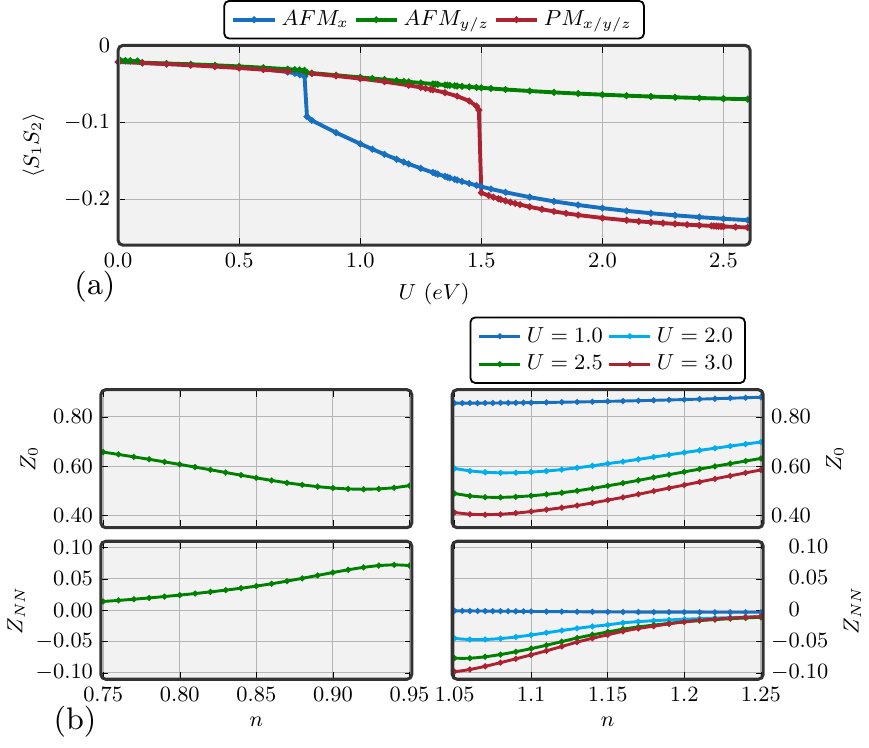}\\
(c)\hspace*{-0.2cm}\includegraphics*[width=8cm]{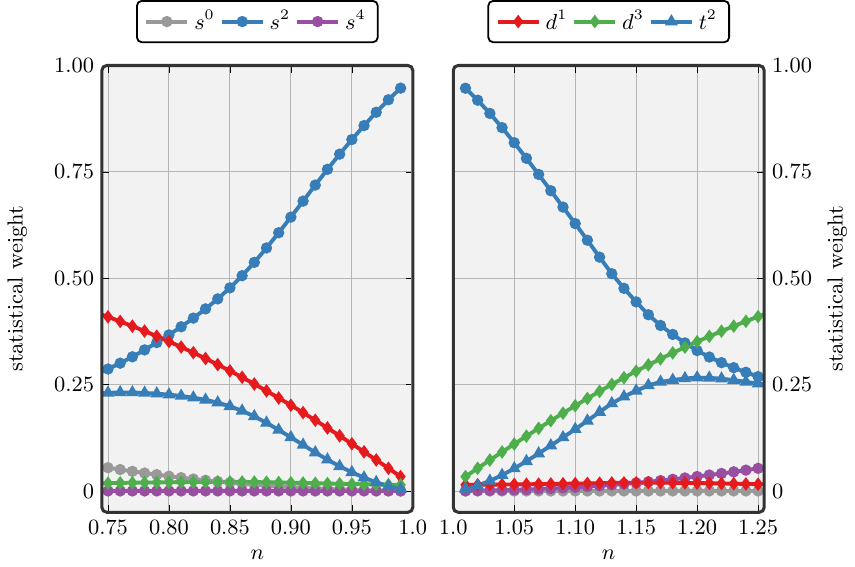}
\caption{(color online) Two-site-cluster observables. 
(a) NN spin-correlation function at half filling. (b) Intra- and inter-site QP 
weight for hole (left) and electron (right) doping from 5-25\%. The NN QP weights 
are positive (negative) for hole (electron) doping. 
(c) Statistics of cluster multiplets with hole (left) and electron (right) doping 
for $U$=2.5eV. Circles denote (s)inglet states, diamonds (d)oublets and triangles 
(t)riplets. The particle sectors are color encoded and marked by the superscript 
numbers. 
\label{fig:multiplets}}
\end{figure}
\begin{figure}[t]
\begin{center}
(a)\includegraphics*{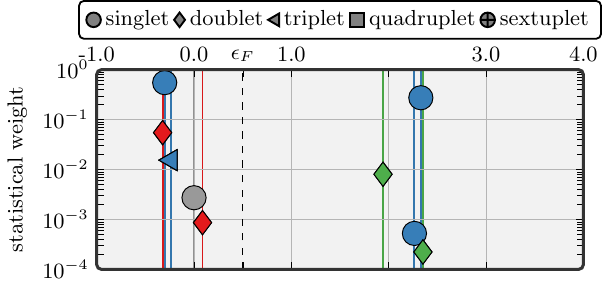}
(b)\includegraphics*{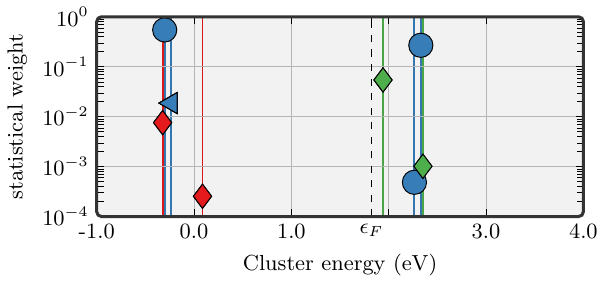}
\end{center}
\caption{(color online) Two-site cluster spectra for $U$=2.5eV with filling 
(a) n=0.95 and (b) n=1.05, which amounts to 5\% hole/electron doping. Colors and
symbols mark states as described in Fig.~\ref{fig:multiplets}c.
\label{fig:spectra-2site-fp}}
\end{figure}
Already the minimal in-plane two-site cluster involving NN Ir sites 
(cf. Fig.~\ref{fig:intro}d) allows for insights on the key effects of an inter-site 
self-energy $\Sigma_{12}$. At half filling, the PM Mott transition occurs at 
$U^{\rm (2)}_{\rm c,PM}$=1.5eV, accompanied by a jump of the 
already negative NN spin-correlation $\langle S_1 S_2\rangle$ towards even lower 
values (cf. Fig.~\ref{fig:multiplets}a). This marks the dominance of the inter-site 
singlet cluster state in the Mott-insulating regime (see below). When allowing for 
the AFM phase, Fig.~\ref{fig:multiplets}a displays that the MIT occurs as in the 
two-single-site study at $U^{\rm (2)}_{\rm c,AFM}$=$U_{\rm c,AFM}$=0.8eV. 

In the doped cases, we focus on cluster effects in the PM phase. 
Figures~\ref{fig:multiplets}b-c show key information on the significance of 
non-local self-energy terms for electron and hole doping. The on-site QP weight
is lower in the electron-doped case for the same value of $U$=2.5eV, marking somewhat
stronger electron correlations. Inter-site (NN) QP weights become relevant for 
$U$$>$1eV. Their magnitude is sizable at small doping and negligible about 20\% away 
from half filling. Note the sign change of $Z_{\rm NN}$ when going from hole to
electron doping.
For sizable $U$ the two-particle singlet on the two-site cluster dominates the
multiplet states at half filling ($n$=1). With doping, increasing weight is 
transferred to the triplet as well as one(three)-particle states when adding 
holes(electrons). Also here there is a small electron-hole asymmetry: 
the singlet(triplet) is more(less) pronounced with hole- than electron doping.
Figure~\ref{fig:spectra-2site-fp} shows for illustration the two-site cluster
spectrum of relevant multipltes with the interacting Fermi level $\varepsilon_F$
for 5\% electron and hole doping, respectively. The multiplets form roughly two groups 
in energy, split by the interaction $U$=2.5eV, understood from the involvment of 
doubly occupied sites in the higher energy group of states.~\cite{fer09_2} In the 
electron-doped case the multiplets are closer to $\varepsilon_F$ in energy, 
reminiscent of the simple picturing of doping into the upper Hubbard band.

\subsection{Four-site cluster}
The four-site cluster is the proper minimal motive on the correlated square lattice
and it is adequate to account for $d_{x^2-y^2}$-ordering tendencies in hole-doped 
cuprates.~\cite{lic00} We utilize it here to include NNN self-energy effects for 
doped iridates in the PM phase. The paramagnetic Mott transition at half filling is 
located at $U^{\rm (4)}_{\rm c,PM}$=1.95eV, correcting for the too dominant NN singlet
formation in the two-site cluster approach.
\begin{figure}[t]
\includegraphics*[width=8.5cm]{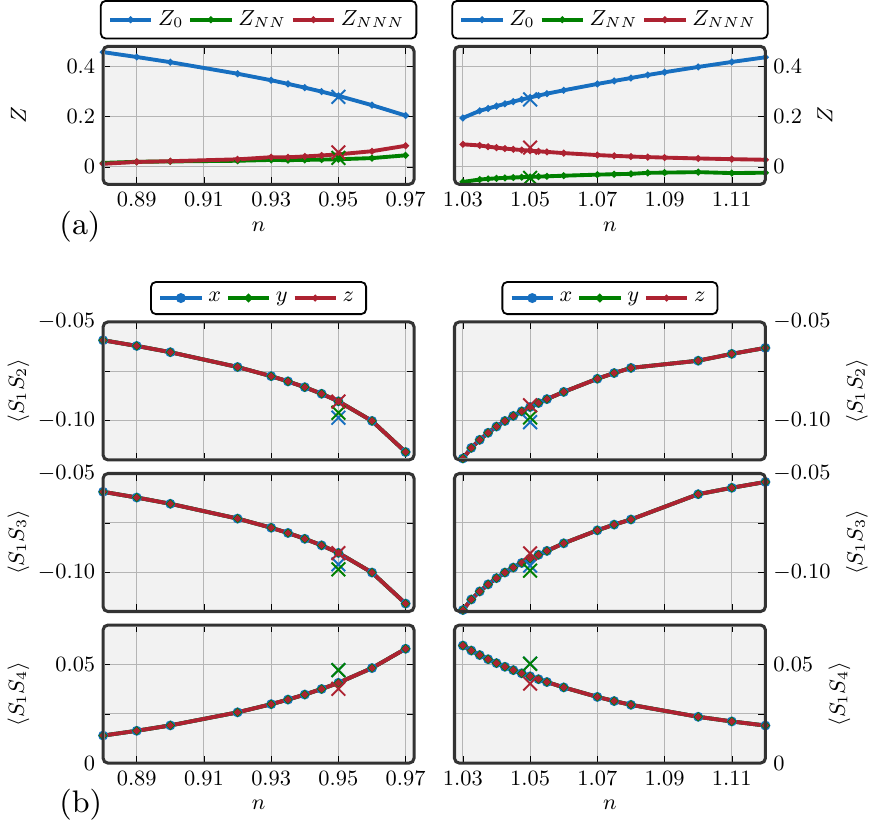}\\[0.1cm]
(c)\hspace*{-0.3cm}\includegraphics*[width=8.5cm]{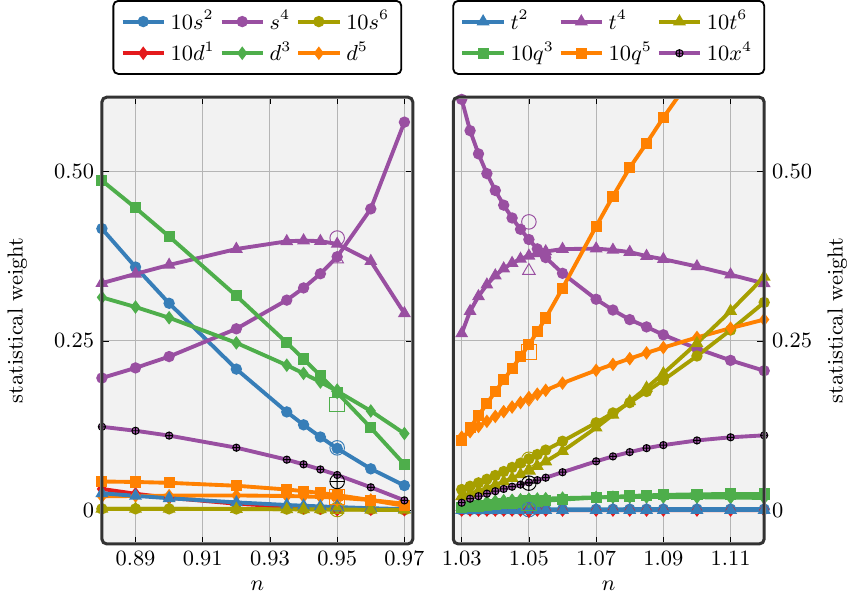}
\caption{(color online) Four-site cluster observables for $U$=2.5eV. 
Left panels show hole doping, right ones electron doping.
(a) QP weights without (full lines) and with inclusiong of the pseudo-spin 
anisotropy term (large crosses). (b) NN and NNN pseudo-spin correlation functions 
$\langle S_iS_j\rangle$.
(c) Statistics of cluster multiplets with hole (left) and electron (right) doping.  
The prefactor $10$ denotes statistical 
weight multiplied by ten for better visibility. Circles denote (s)inglets, 
diamonds (d)oublets, triangles (t)riplets, squares (q)uartets and crossed circles 
sextuplets (x). Open symbols at 5\% doping represent the matching states from
inclusion of the pseudo-spin anisotropy.}
\label{fig:four-site}
\end{figure}

In the four-site cluster description the correlation strength for the same value
of $U$ is generally enhanced compared to the two-site cluster approach, documented 
by the smaller on-site QP weight in Fig.~\ref{fig:four-site}a. Moreover the relation 
$|Z_{\rm NNN}|$$>$$|Z_{\rm NN}|$ holds for small doping, pointing towards anisotropic
electron correlations in this regime. A larger correlation anisotropy is expected 
in the electron-doped compound because of the sign difference between $Z_{\rm NNN}$ 
and $Z_{\rm NN}$. For any given symmetric doping the on-site $Z$ is marginally
lower for electron doping. The pseudo-spin anisotropy renders the calculations
numerically more challenging and is thus only included at 5\% doping. There it 
leads again to a marginal increase of correlation strength.
The pseudo-spin correlation function $\langle S_iS_j\rangle$ on the four-site
cluster depicted in Fig.~\ref{fig:four-site}b has strong AFM signature in NN 
distance and conclusively strong FM signature in NNN distance, both monotonically
decreasing from half filling. With symmetric doping the respective pseudo-spin 
correlations are somewhat stronger in the electron-doped case. As expected, including 
the anisotropy term in the Hamiltonian strengthens the in-plane correlations,
especially alongside the commensurate directions, i.e. along $x$ for 
$\langle S_1S_2\rangle$ and along $y$ for $\langle S_1S_3\rangle$. Though energetically
no in-plane easy-axis is favored, AFM order is numerically most easily stabilized with
the experimental [110] easy-axis. 
\begin{figure}[t]
\begin{center}
(a)\includegraphics*{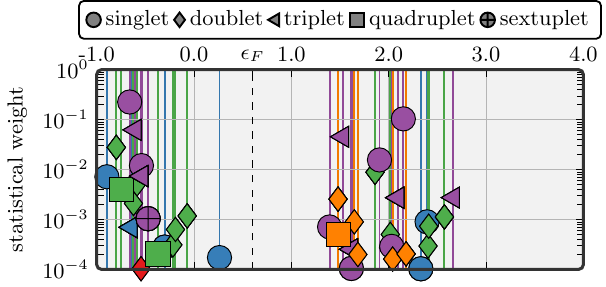}
(b)\includegraphics*{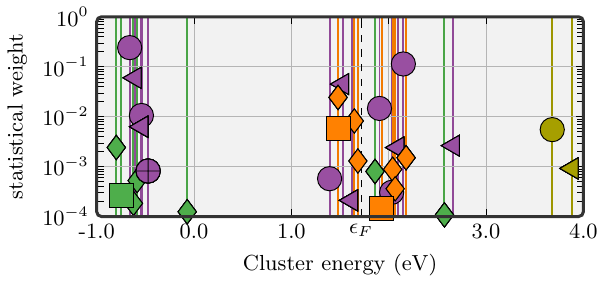}
\end{center}
\caption{(color online) Four-site cluster spectra without pseudo-spin anisotropy
for $U$=2.5eV. 
(a) n=0.95 (b) n=1.05, which amounts to five percent doping, respectively. Colors and
symbols mark states as described in Fig.~\ref{fig:four-site}c.}
\label{fig:spectra-4site}
\end{figure}

Figure~\ref{fig:four-site}c shows the statistical weight of the four-site
cluster multiplets with doping when neglecting the pseudo-spin anisotropy term. 
Dominant singlet states, now in the four-particle sector, rule 
again at small doping, but in contrast to the two-site cluster approach the
triplet states take over beyond 5\% hole or electron doping. Furthermore 
including the NNN self-energy, connected to the NNN hopping, now leads to
marginally stronger(weaker) triplets(singlets) in the hole-doped regime. Thus
short-range spin-flucuations should be slightly larger for hole doping. Charge
fluctuations on the electron-doped side from the four-particle into the five-particle
cluster sector are more pronounced than the symmetric fluctuations on the hole-doped
side from the four-particle into the three-particle sector. Moreover even fluctuations
into the six-particle sector are taking place both into singlet and triplet states.
With hole doping, only the two-particle singlet has some weight while the two-particle
triplet is negligible. Eventual inclusion of the pseudo-spin anisotropy term 
enhances the singlet-triplet splitting in the dominant four-particle sector.
At 5\% doping, the interacting Fermi level is again located in higher energy block of 
multiplets for electron doping, while for hole doping it remains more or less 
inbetween both blocks of multiplets (see Fig.~\ref{fig:spectra-4site}). Thus also
in this larger-cluster approach a doped-Mott-insulator picture applies more to
the electron-doped regime.

In order to assess the electronic correlation strength with hole and electron doping,
still a further viewpoint can be taken. As discussed in previous 
works,~\cite{byc12,thu12} the computation of the local von-Neumann entropy $S$ 
may provide a measure of correlation. The off-diagonal cluster density matrix 
$\rho_{qq'}$ (see eq. (\ref{statweight})) may be used to compute $S$ and 
relative entropies. 
After diagonalising $\rho_{qq'}$, its eigenvalues $\rho_\lambda$ are utilized to 
write the local von-Neumann entropy via $S$=$-\sum_\lambda \rho_\lambda\,
\ln \rho_\lambda$ as well as the relative entropy 
$\Delta S(\rho^A||\rho^B)$=$\sum_\lambda \rho^A_\lambda (\ln \rho^A_\lambda - \ln\rho^B_\lambda )$ for two systems $A$ and $B$. The larger the relative entropy, the more 
distinct the two compared systems are.

\begin{figure}[b]
\centering
\includegraphics*[width=8.5cm]{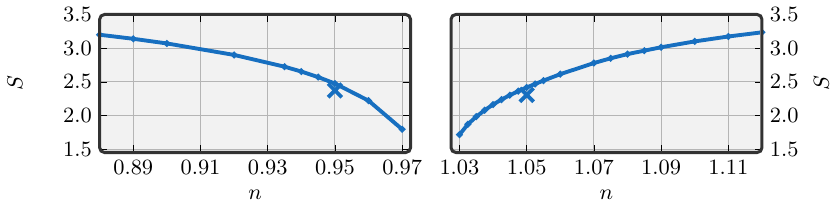}
\caption{(color online) Local von-Neumann entropy on the four-site cluster with 
doping. Crosses: with inclusion of  the pseudo-spin anisotropy.}
\label{fig:4site-entropy}
\vspace*{0.2cm}
\begin{tabular}[c]{|c|c|c|c|c|}
\hline
doping & $S(\rho)$ & $S(\rho^{0})$ & $S(\rho||\rho^{0})$ & $S(\rho^{0}||\rho)$ \\
	\hline
	$-5 \%$ & 2.37 & 4.04 & 1.63 & 0.0 \\
	\hline
	$+5 \%$ & 2.31 & 4.08 & 2.18& 0.0 \\
	\hline
\end{tabular}
\captionof{table}{Local von-Neumann entropy at 5\% symmetric doping including
pseudo-spin anisotropy. The distribution $\rho$ marks the interacting ensemble and
$\rho^0$ is associated with the non-interacting ensemble of states.}
\label{tab:4site-entropy}
\end{figure}

Figure~\ref{fig:4site-entropy} shows that at low symmetric doping the entropy $S$ 
is slightly smaller in the electron-doped case, rendering it more correlated. 
Inclusion of the pseudo-spin anisotropy again enhances the correlation effect. 
Albeit the electron-hole correlation asymmetry from entropy is small in 
absolute numbers, the relative entropy by comparison to the non-interacting case
marks the electron-doped regime rather clearly as the one with increased correlation
strength (see Tab.~\ref{tab:4site-entropy}).

\begin{figure}[t]
\begin{center}
(a)\includegraphics*[width=3.8cm]{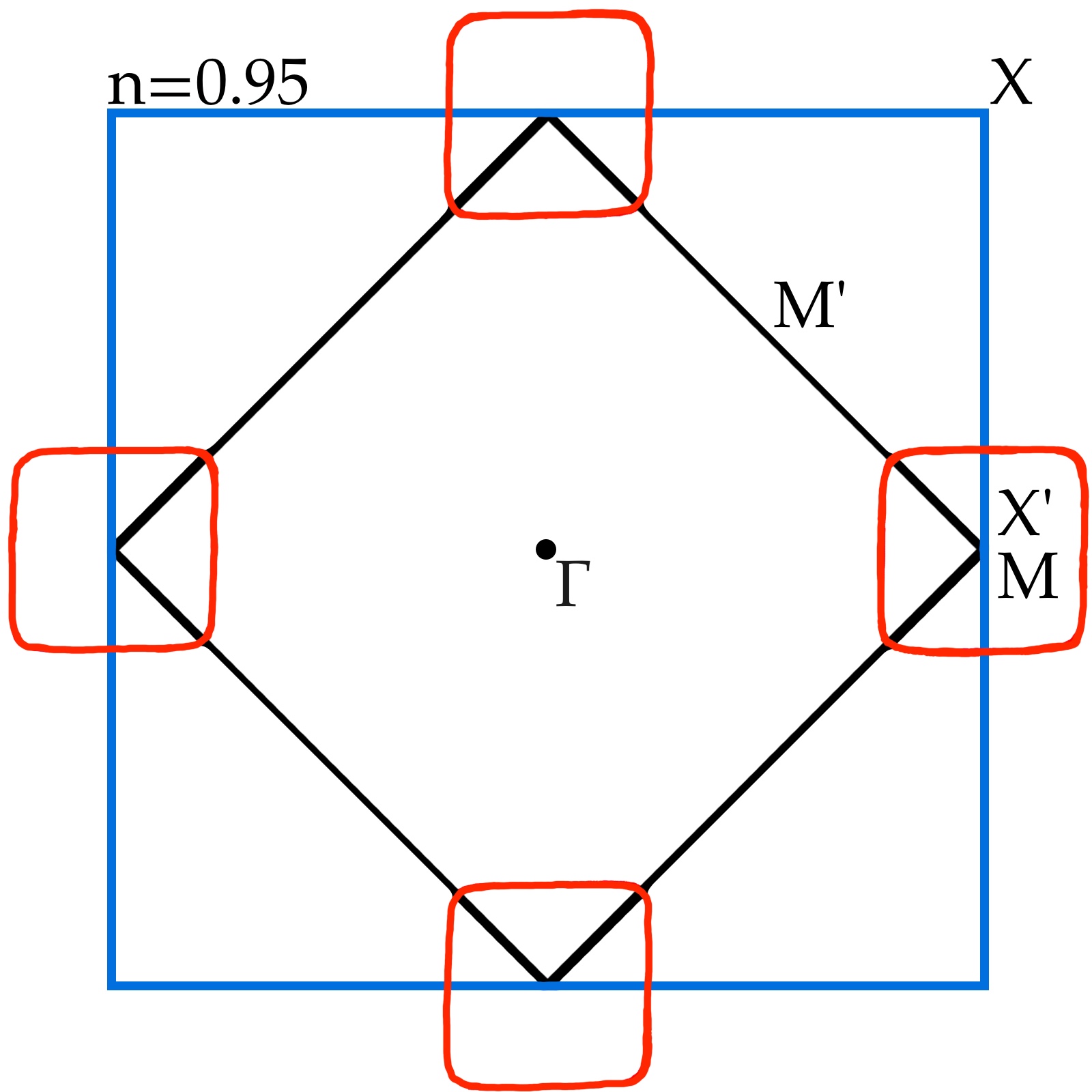}
(b)\includegraphics*[width=3.8cm]{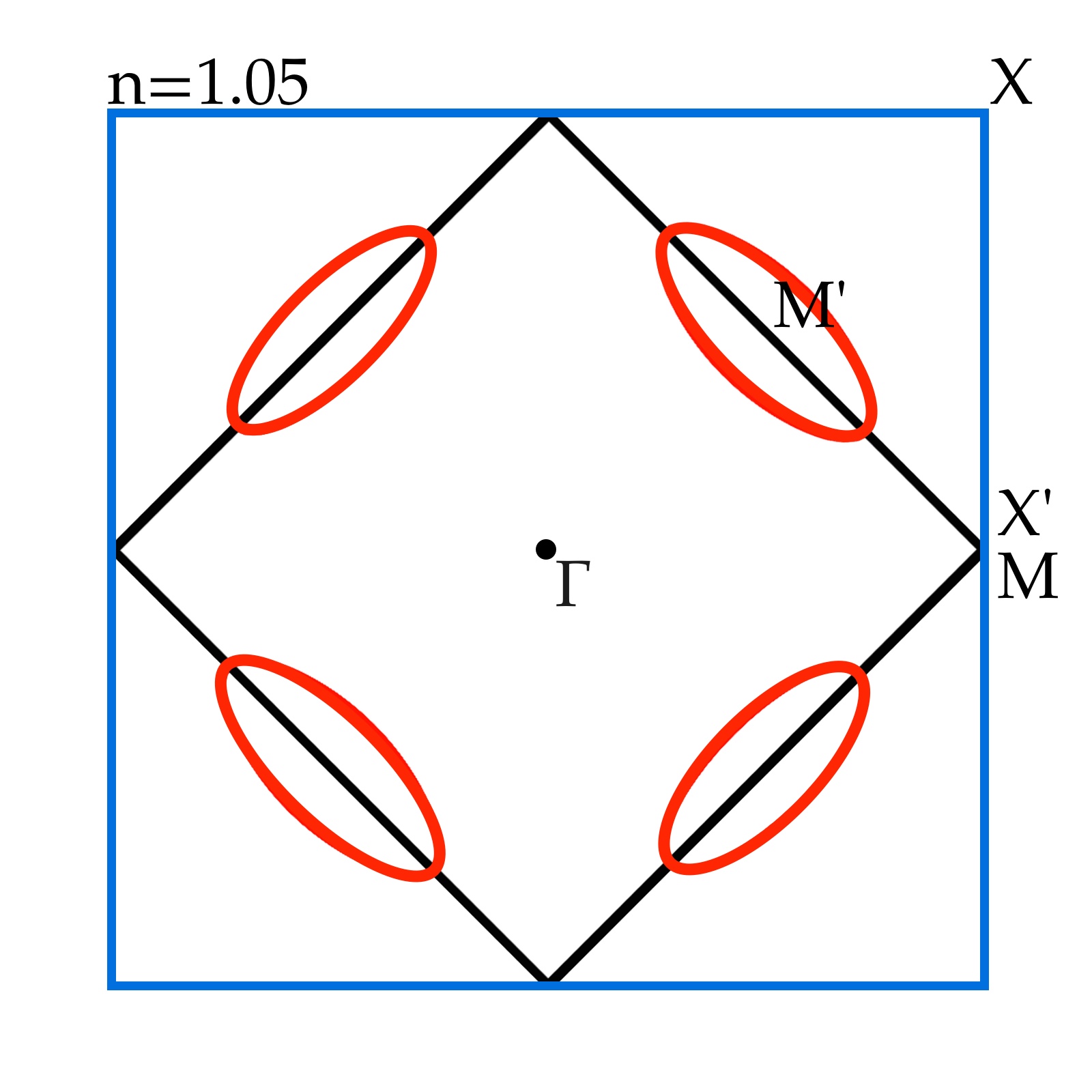}\\[0.1cm]
(c)\includegraphics*[width=3.8cm]{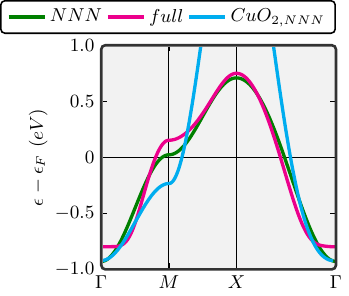}
(d)\includegraphics*[width=3.8cm]{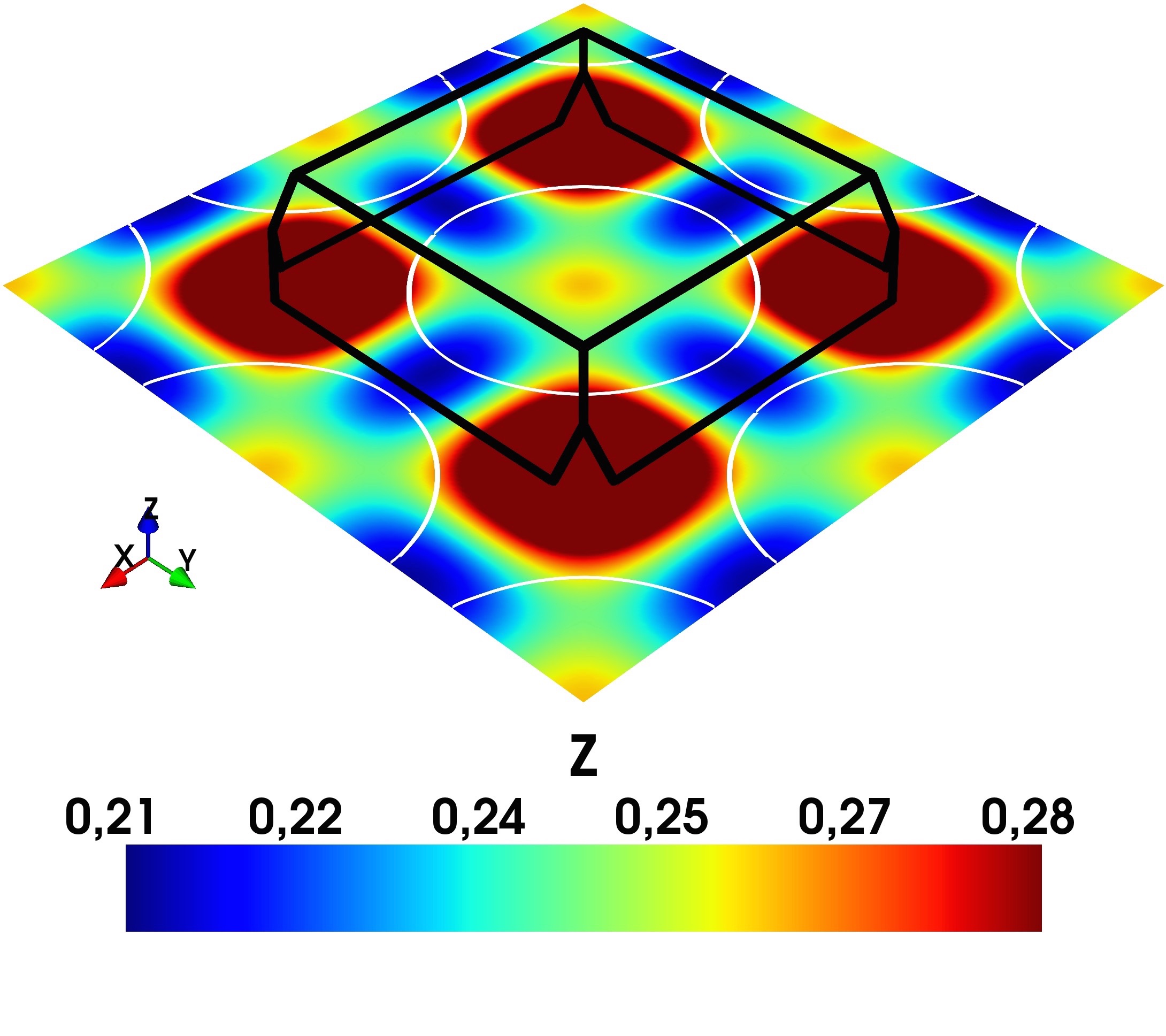}\\[0.1cm]
(e)\includegraphics*[width=3.8cm]{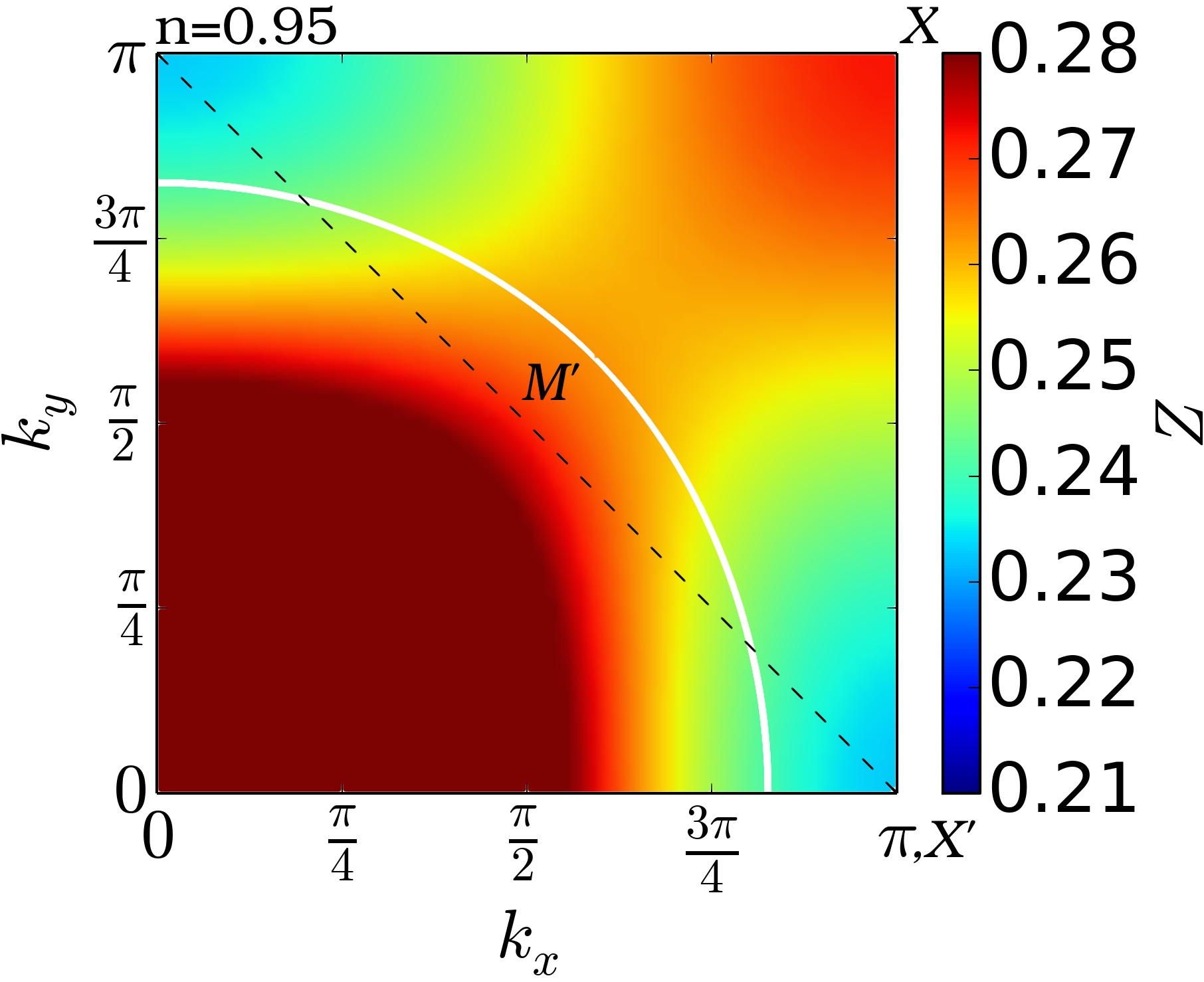}
(f)\includegraphics*[width=3.8cm]{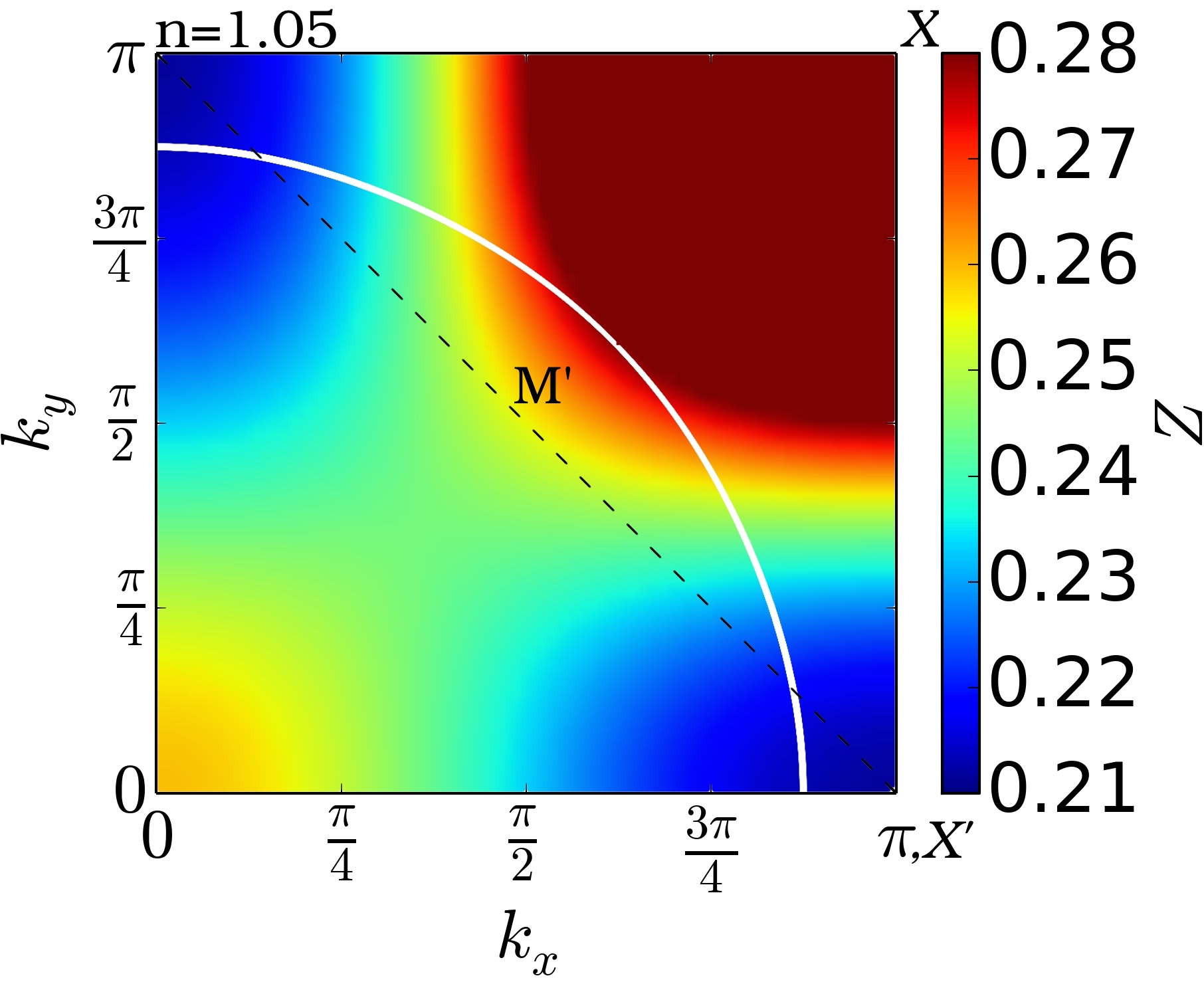}
\end{center}
\caption{(color online) Iridate $k$-selectivity.
(a,b) FS pockets from the two-single-site treatment for hole and electron doping
at $U$=1eV. (c) NNN- and full-hoppings one-band dispersion compared with
standard cuprate dispersion (CuO$_2$: $t$=$-$430meV, $t'$=$+$129meV~\cite{pav01}).
(d) in-plane QP weight $Z(\bk)$ for $U$=2.5eV and electron doping $n$=1.05, 
black line: Brillouin zone, white line: interacting FS.
(e,f) Magnified comparison in symmetry-inequivalent $k$-space sector, between
(e) hole doping and (f) electron doping, both for $U$=2.5eV.\label{fig:kdep}}
\end{figure}

Finally we want to discuss $k$-dependent signatures at finite doping based on the
four-site cluster approach. In principle 
two scenarios may hold: either doping right within the Slater-Hubbard bands takes 
place ($U$$\sim$$W$), or it results in the build-up of a renormalized FS readily from 
the original itinerant dispersion ($U$$\gg$$W$). In the first case, $k$-space 
differentiation occurs because of the energy dependence of the gap-forming bands 
(compare Fig.~\ref{fig:afmmit}b). Then here, hole(electron) doping would lead to 
FS pocket-formation around $X'$($M'$), as indeed verified by plotting the doped FS 
within our two-single-site approach in Fig.~\ref{fig:kdep}a,b. Such a scenario 
apparently has been detected in ARPES measurements for effective {\sl hole doping} of 
Sr$_2$IrO$_4$.~\cite{cao14}

In the strong-coupling scenario, $k$-selectivity in a one-band picture is 
usually due to finite inter-site terms $\Sigma_{\alpha\beta}$$\ne$0 for 
$\alpha$$\ne$$\beta$. We may encounter such effects via our periodized in-plane 
cluster self-energies. For instance, the QP weight $Z$=$Z(\bk)$ for the  
model (\ref{eq:1bham}) can vary based on NN and NNN self-energies
of the four-site cluster. Figures~\ref{fig:kdep}d-f display the obtained QP variations
in the Brillouin zone without pseudo-spin anisotropy. As discussed before, inclusion
of the latter generally leads to a minor increase of the overall correlation strength.
Lets focus on the interacting fermiology, i.e., $Z$=$Z(\bk_F)$, where $\bk_F$ is 
the Fermi wave vector. For both dopings, i.e. hole- and electron-like, 
Figs.~\ref{fig:kdep}e,f show an obvious in-plane nodal/anti-nodal dichotomy. The QP 
weight on the FS along the node $(0,0)$$-$$(\nicefrac{\pi}{2},\nicefrac{\pi}{2})$ is 
larger than along the anti-node $(0,0)$$-$$(0,\pi)$. Though the absolute differences
are small within cluster-RISB, it serves as a proof of principles for $k$-space 
differentiation by electronic correlations, in agreement with recent ARPES 
experiments on surface {\sl electron-doped} Sr$_2$IrO$_4$.~\cite{kim14} 

Second, there is a substantial quantitative difference in the $k$-space differentiation of 
$Z(\bk_F)$ between both doping directions. The electron-doped case exhibits stronger
QP-weight variation along $\epsilon_{\rm F}$ than the hole-doped case. In other words 
for same interaction strength, theory predicts that electron doping of single-layer 
iridates is more likely to cause a Fermi-arc structure than hole doping. This 
finding is reminiscent of the electron-hole dichotomy in cuprates,~\cite{arm10,web10} 
yet with a twist: in cuprates, the hole-doped case is assumed more susceptible to
$k$-selective correlations. Generally, for all encountered symmetric doping 
distances from $n$=1, the intra-site $Z$ is always somewhat lower on the 
electron-doped side.

As pointed out before,~\cite{wat10,wat13} the qualitative difference may be 
explained by the relevance of hopping characteristics beyond NN.~\cite{pat09}
Because of the different sign of the NNN $t'$ in both compound families, the enhanced 
correlation-susceptible van-Hove singularity at $M$ in reciprocal space 
is above(below) the Fermi level for iridates(cuprates) as shown in 
Fig.~\ref{fig:kdep}c. Thus from a phase-space argument, hitting stronger correlations
at the anti-node takes place by electron(hole) doping of iridates(cuprates). The
hoppings beyond NNN are then effective in shifting the iridate van-Hove 
singularity further away from $\epsilon_{\rm F}$. Note that a recent extended
fluctuation-exchange-based study~\cite{li15} also found electron-hole doping 
asymmetries in Sr$_2$IrO$_4$.

\section{Summary}
An effective $J$=$\nicefrac{1}{2}$ low-energy one-band modelling is derived for
single-layer iridates from the initial spin-orbit interacting $t_{2g}$ manifold. 
For $U$$\lesssim$1.25eV Slater-like behavior dominates, while for $U$$\gtrsim$1.25eV 
Mott-Hubbard physics is more in control. In reality a subtle interplay between both 
limits is expected.~\cite{hsi12} Our theoretical study reveals an 
electron-hole doping asymmetry approached from two directions. First the
analysis of QP weights and local cluster states at strong coupling points to 
increased electronic correlations on the electron-doped side. Second, investigating 
the low-energy $k$-space differentiation also exposes a doping asymmetry, taking 
place at weaker as well as at stronger coupling and has partly already been 
confirmed by recent experiments.~\cite{kim14,cao14} Fermi-surface
pockets that occur for weaker electron-electron interaction are more likely for hole 
doping, whereas Fermi arcs may set in for stronger interaction with higher tendency 
again on the electron-doped side. Therefore electron-doped iridates are candidates 
for a possible analogue to hole-doped cuprates. The inclusion of the small pseudo-spin
anisotropy is shown to somewhat increase the correlation strength, but no 
drastic qualitative changes arise therefrom at the present level of modeling.

\begin{acknowledgments}
We are indebted to F. Baumberger, M. Behrmann, L. Boehnke, D. S. Dessau, R. Heid 
and A. I. Lichtenstein for helpful discussions. This research was supported by the 
DFG-FOR1346 project.
Computations were performed at the North-German Supercomputing Alliance (HLRN) 
under Grant No. hhp00031. 
\end{acknowledgments}

\bibliographystyle{apsrev4-1}
\bibliography{bibextra}

\end{document}